# Logical Classification of Partially Ordered Data[*]

Elena V. Djukova [1], Gleb O. Masliakov [2], Petr A. Prokofyev[3]

[1] Federal Research Center "Computer Science and Control,"
Russian Academy of Sciences, Moscow, 119333 Russia
`edjukova@mail.ru`
[2] Moscow State University, Moscow, 119899 Russia
`gleb-mas@mail.ru`
[3] Mechanical Engineering Research Institute,
Russian Academy of Sciences, 101000, Moscow, Russia
`p_prok@mail.ru`

**Abstract.** Issues concerning intelligent data analysis occurring in machine learning are investigated. A scheme for synthesizing correct supervised classification procedures is proposed. These procedures are focused on specifying partial order relations on sets of feature values; they are based on a generalization of the classical concepts of logical classification. It is shown that learning the correct logical classifier requires an intractable discrete problem to be solved. This is the dualization problem over products of partially ordered sets. The matrix formulation of this problem is given. The effectiveness of the proposed approach to the supervised classification problem is illustrated on model and real-life data.

## 1 Introduction

In classification problems, the training data is a set of examples of objects under examination in which each object is represented by a numerical vector obtained by measuring or observing its parameters. The properties of objects to be measured or observed are called features. In a simple case, the examples are divided into two classes—the class of positive and the class of negative examples. In the general case, the number of classes may be greater than two. Given a description of an unknown object in terms of features, it is required to find out (recognize) the class it belongs to.

The main advantage of the logical approach to the classification (recognition) problem is the possibility to obtain a results without additional probabilistic assumptions and using a small number of training objects (using a small number of precedents). The analysis of training data is reduced to finding certain dependences or ele-mentary classifiers (which are subsets of feasible values of some features) that differentiate objects belonging to different classes. An object is classified judging by the presence or absence of such elementary classifiers in the object's description. Special attention is given to synthesizing correct algorithms, i.e., algorithms that unmistakably classify the training objects.

[*] This study was partially supported by the Russian Foundation for Basic Research, project no. 19-01-00430-a.

Models of correct logical classifiers based on finding "correct" elementary classifiers are most effective in the case of integer data with a small number of possible values, especially binary data. Examples are classification by the vote of tests or by the vote of representative sets or by the vote of class coverings. The first models of such classifiers were proposed in [Baskakova et al., 1981] and [Dmitriev et al., 1966], and their description using the concept of elementary classifier was first proposed in [Djukova et al., 2002].

There are complicated problems in which no sufficient number of informative correct elementary classifiers can be found. For example, such a situation occurs when the features can take a large number of possible values. Features that can take real values are often treated as integer valued features with a large number of possible values. A way to solve such problems is to use logical correctors. Here we mean correct recognition algorithms based on constructing correct sets elementary classifiers of from incorrect elementary classifiers [Djukova et al., 2017a] and [Djukova et al., 1996].

In [Djukova et al., 2017a], a generic scheme for synthesizing correct logical classification procedures was proposed; within this scheme, classical models and logical correctors were described.

If the feature space is large, computationally complex (intractable) problems have to be solved. The central place among these problems is occupied by the monotone dualization problem, i.e., the problem of constructing a reduced disjunctive normal form of a monotone Boolean function specified by a conjunctive normal form. The intractability of the monotone dualization problem has two aspects—an exponential growth of the number of solutions as the problem size increases and the complexity of finding (enumerating) these solutions. The most efficient algorithms are algorithms with a polynomial step (with a polynomial delay) However, polynomial algorithms are known only for some special cases of monotone dualization (e.g., see [Johnson et al., 1988]).

Application problems cannot always be described within the classical statement of logical classification in which feature values are compared for equality. In many classification problems, each feature can take values from a partially ordered set.

In this paper, we propose a scheme for synthesizing correct logical algorithms under the condition that partial order relations are specified on the sets of values of integer-valued features. The basic concepts used in the logical analysis of integer data in the supervised classification problem are generalized, and conditions for the correctness of the basic logical classification procedures are obtained. It is found that the analysis of training samples with partial orders requires the dualization problem over the product of finite partial orders to be solved; a simple special case of this problem is monotone dualization. We give a matrix formulation of the general dualization problem and show that this problem is reduced to enumerating special coverings of an integer matrix, which are called in this paper ordered irredundant coverings. The concept of the ordered irredundant covering of an integer matrix is a generalization of the well-known concept of irreducible covering of a Boolean matrix used in the matrix formulation of the monotone dualization problem. Using real-life data, we establish the dependence of the quality of logical classification on the choice of partial orders on the sets of feature values.

## 2   Logical Classification of Integer Data in the Classical Statement

Consider the supervised classification problem. Let $M$ be the set of objects under examination; this set is divided into $l$ classes $K_1, \ldots, K_l$. Let the objects in $M$ be described by the features $x_1, \ldots, x_n$. Let us define the basic concepts used in designing the classical logical classification procedures. We assume that the set of values of each feature consists of integer numbers and this set is finite.

Let $H = \{x_{j_1}, \ldots, x_{j_r}\}$ be a set of $r$ different features, and $\sigma = (\sigma_1, \ldots, \sigma_r)$, where $\sigma_i$ is a possible value of the feature $x_{j_i}$, $i = 1, 2, \ldots, r$. The pair $(\sigma, H)$ is called an elementary classifier (el.cl.) of rank $r$. The proximity of the object $S = (a_1, \ldots, a_n)$ in $M$ to the el.cl. $(\sigma, H)$ is determined by the quantity $B(\sigma, S, H)$, which is equal to 1 if $a_{j_t} = \sigma_t$ for $t = 1, 2, \ldots, r$ and is equal to 0, otherwise. If $B(\sigma, S, H) = 1$, then the object $S$ is said to generate (contain) the el.cl. $(\sigma, H)$. The el.cl. $(\sigma, H)$ is correct for the class $K$ ($K \in \{K_1, \ldots, K_l\}$) if no pair of training objects $S'$ and $S''$ exists such that $S' \in K$, $S'' \notin K$, and $B(\sigma, S', H) = B(\sigma, S'', H) = 1$.

The classification algorithm $A$ in the phase of learning constructs for each class $K$ a set of correct el.cl. $C^A(K)$. The classification of an object $S$ is based on computing the quantity $B(\sigma, S, H)$ for each el.cl. $(\sigma, H)$; i.e., each element of the set $C^A(K)$ takes part in the voting procedure. As a result, an estimate of the membership of the object $S$ in the class $K$ is found. Let us consider the basic models.

In the general case, the elementary classifier $(\sigma, H)$ can possess one of the following two properties with respect to the class $K$: (1) some training objects in $K$ contain $(\sigma, H)$; (2) none of the training objects in $K$ contains $(\sigma, H)$.

Every correct el.cl. of the first type is called a representative el.cl. for the class $K$. A representative el.cl. $(\sigma, H)$ for the class $K$ is said to be an irredundant representative el.cl. of the class $K$ if any el.cl. $(\sigma', H')$ such that $\sigma' = (\sigma_1, \ldots, \sigma_{t-1}, \sigma_{t+1}, \ldots, \sigma_r)$ and $H' = H \setminus \{x_{j_t}\}$ for $t \in \{1, 2, \ldots, r\}$ is not a representative el.cl. for $K$.

A set of features $H$ is called a test if each precedent of each class $K$ ($K \in \{K_1, \ldots, K_l\}$) contains a representative el.cl. for $K$ of the form $(\sigma, H)$. A test is called irredundant if its every proper subset is not a test.

The elementary classifier of the second type is called a covering of the class $K$. A covering $(\sigma, H)$ of $K$ is said to be an irredundant covering of $K$ if any el.cl. $(\sigma', H')$ such that $\sigma' = (\sigma_1, \ldots, \sigma_{t-1}, \sigma_{t+1}, \ldots, \sigma_r)$ and $H' = H \setminus \{x_{j_t}\}$ for $t \in \{1, 2, \ldots, r\}$ is not a covering of $K$.

In simple modifications of classification algorithms based on construction of a set of representative el.cl., the estimate for the class $K$ is computed by summing

4the quantities $P_{(\sigma,H)}B(\sigma,S,H)$, where $(\sigma,H) \in C^A(K)$ and $P_{(\sigma,H)}$ is the number of training objects in K containing $(\sigma,H)$. In the classification algorithms based on construction of a set of class coverings, the estimate for the class K is computed by summing the quantities $1 - B(\sigma,S,H)$ over $(\sigma,H) \in C^A(K)$.

Elementary classifiers of small rank are most informative. For this reason, in applications the rank of el.cl. is limited or only irredundant correct el.cl. are considered (and even not all of them). However, in problems in which features have a large number of possible values, almost all correct elementary classifiers have a large rank, and consequently all such classifiers have low informativeness. Problems in which features have too large number of possible values are too complicated for the logical classifiers considered above; in this situation, as has been mentioned in the Introduction, the use of logical correctors is more effective.

## 3  Logical Data Analysis with Partial Orders

In this section, we give the basic definitions of logical data analysis with partial order relations and formulate two key problems of this analysis—the problem of finding the maximal independent elements of the product of partial orders and the problem of finding the minimal independent elements of the product of partial orders.

Let $P = P_1 \times \ldots \times P_n$, where $P_1, \ldots, P_n$ are finite partially ordered sets. An element $y = (y_1, \ldots, y_n) \in P$ follows an element $x = (x_1, \ldots, x_n) \in P$ ($x$ precedes $y$) if $y_i$ follows $x_i$ ($x_i$ precedes $y_i$) for $i = 1, 2, \ldots, n$. We use the notation $x \leqslant y$ to denote that $y \in P$ follows $x \in P$ ($x$ precedes ). The notation $x \prec y$ means that $x \leqslant y$ and $y \neq x$. The elements $x, y \in P$ are called comparable if $x \leqslant y$ or $y \leqslant x$. Otherwise, $x$ and $y$ are said to be incomparable. The greatest element in $P$ is the element $x$ for which $y \leqslant x$ for every y $\in P$.

Let $R \subseteq P$. We introduce the following notation: $R^+ = R \cup \{x \in P | \exists a \in R, a \prec x\}$ is the set of elements following the elements of $R$; $R^- = R \cup \{x \in P | \exists a \in R, x \prec a\}$ is the set of elements preceding the elements of $R$. The element $x$ in $P\setminus R^+$ ($P\setminus R^-$) is called the maximal (minimal) element of $P$ independent of $R$ if, for any other element $y$ in $P\setminus R^+$ ($P\setminus R^-$), the relation $x \prec y$ ($y \prec x$) does not hold.

Denote by $I(R^+)$ the set consisting of the maximal independent of $R$ elements of the set $P$, and denote by $I(R^-)$ the set consisting of the minimal independent of $R$ elements of $P$. For the given set $R$, we want to construct the sets $I(R^+)$ and $I(R^-)$. The sets $I(R^+)$ and $I(R^-)$ are dual of each other: if $Q = I(R^-)$ or $Q = I(R^+)$, then, respectively, $I(Q^+) = I(R^+)$ and $I(Q^-) = I(R^-)$. In what follows, each of these problems is called the dualization problem over the product of partial orders.

One of the most popular and well-studied case is the one in which the sets $P_1, \ldots, P_n$ are chains, i.e., any two elements in these sets are comparable. An antichain is a set in which any two elements are incomparable. The sets $I(R^+)$ and $I(R^-)$ are antichains.

The simplest case of dualization over the product of chains is monotone dualization. The problem is formulated as follows.

Let a conjunctive normal form realizing the monotone Boolean tion $F(x_1, \ldots, x_n)$ be given. We want to construct a reduced disjunctive normal form of $F$. It is easy to verify that monotone dualization is equivalent to the problem of constructing the set $I(R^-)$ under the condition that $P$ is the $n$-dimensional Boolean cube, $P_i = \{0, 1\}$ for $i = 1, 2, \ldots, n$, the order $0 \prec 1$ is defined in each $P_i$, and $R$ is the set of zeros of the function $F$ containing the set of upper zeros of $F$. The elements of $I(R^-)$ are lower units of the function.

The task of monotone dualization can be formulated in terms of hypergraphs [Murakami et al., 2014] and in terms of matrices using the concept of irreducible covering of a Boolean matrix [Djukova et al., 2015].

Theoretical estimates of the effectiveness of dualization algorithms are based on evaluating the complexity of a single step, i.e. the complexity of finding a new solution. In the most efficient algorithm, the step complexity is polynomial in the input size. Algorithm of this kind is called a polynomial-delay algorithm. However, polynomial algorithms have been constructed only for some particular cases of monotone dualization, for example, for the case of 2-CNF [Johnson et al., 1988]. Currently, there are two main areas of research.

The first direction aims to construct so-called incremental algorithms, when the algorithm is allowed to review the solutions obtained in previous steps. The complexity of the algorithm step is estimated for the worst case (for the most complex variant of the problem). In [Friedman et al., 1996] an algorithm of monotone dualization with quasipolynomials step determined by the size of both its input and its output has been constructed.

In [Boros et al., 2002] for the case where each $P_i$, $i = 1, 2, \ldots, n$, is a chain and $|P_i| \geq 2$, a quasi-polynomial incremental algorithm based on the algorithm proposed in [Fredman et al., 1996] is constructed. A similar result is obtained for some other special finite partial orders (semi-lattices of bounded width, lattices defined by a set of real closed intervals, forests [Elbassioni et al., 2009]). For lattices in [Babin et al., 2017], the impossibility of constructing a polynomial incremental algorithm (otherwise $P = NP$) is proved, and for the case of distributive lattices a subexponential incremental algorithm (output subexponential algorithm) is proposed. It should be noted that the incremental approach is mainly of theoretical interest because in the worst case the number of dualization solutions (the output size) increases exponentially with the increase of the input size.

The second direction is based on the construction of asymptotically optimal dualization algorithms [Djukova, 1977], [Djukova et al., 2015]. In this case, the algorithm is allowed to execute additional polynomial steps on condition that their number is almost always sufficiently small compared with the number of all solutions of the problem. This has led to the construction of algorithms for monotone dualization that are efficient in the typical case (efficient for almost all variants of the problem). These algorithms have a theoretical basis and are leaders by the computational time.

In [Djukova et al., 2017b and Djukova et al., 2018], an asymptotically optimal algorithm RUNC-M+ for the dualization over the product of chains was constructed. The proof of the asymptotic optimality of RUNC-M+ is based on proving the asymptotic equality of two quantities—the typical number of steps of this algorithm and the typical number of solutions to the problem (the cardinality of $I(R^+)$). To this end, a matrix formulation of the problem of constructing $I(R^+)$ for the case of product of chains was given, and the technique of obtaining asymptotic estimates that were earlier used for proving the optimality of monotone dualization algorithms was elaborated. Below, we give a matrix formulation of the problem of constructing the maximal independent elements for the product of finite partial orders P based on a more general concept of ordered irredundant covering of a matrix the rows of which are sets from $R \subseteq P$ than the concept used in [Djukova et. al., 2018] for the case of chains.

Let us introduce the following notation. $Q_1(x, P)$ ($x \in P$) is the set of all elements in P that immediately follow $x$ ($Q_1(x, P) = \{y \in P: x \prec y, \forall a \in P : x \prec a \Rightarrow a \not\prec y\}$); $Q_2(x, y, P)$ ($x \in P$, $y \in Q_1(x, P)$) is the set of all elements a $\in$ P that do not precede x and precede $y$ ($Q_2(x, y, P) = \{a \in P: a \not\preceq x, a \preceq y\}$).

We also define the ordered irredundant covering of the matrix $L_R$ the rows of which are sets from $R \subseteq P$.

Let $H$ be the set of columns of the matrix $L_R$ with indexes $j_1, \ldots, j_r$ and $\sigma = (\sigma_1, \ldots, \sigma_r)$ where $\sigma_i \in P_{j_i}$ for $i = 1, 2, \ldots, r$. Then $H$ is said to be an ordered irredundant $\sigma$-covering if the following two conditions are satisfied: (1) for every $i \in \{1, 2, \ldots, r\}$ and every $y \in Q_1(\sigma_i, P_{j_i})$, the submatrix $L_R^H$ of $L_R$ formed by columns of $H$ contains each row of the form $(\beta_1, \ldots, \beta_{i-1}, \beta_i, \beta_{i+1}, \ldots, \beta_r)$ where $\beta_i \in Q_2(\sigma_i, y, P_{j_i})$ and $\beta_t \preceq \sigma_t$ for $t \neq i$, $t \in \{1, 2, \ldots, r\}$; (2) the submatrix $L_R^H$ does not contain rows preceding $\sigma$.

Note that if $P_i$ ($i \in \{1, 2, \ldots, n\}$) is a finite chain and $x \in P_i$ is not the greatest element in $P_i$, then the set $Q_1(x, P_i)$ consists of a single element, which is denoted by $x + 1$ below; therefore $Q_2(x, x + 1, P_i) = \{x + 1\}$. For this reason, in the case of the product of finite chains, condition (1) in the definition of the ordered irredundant $\sigma$-covering turns into the following condition: for every $i \in \{1, 2, \ldots, r\}$, the submatrix $L_R^H$ of the matrix $L_R$ formed by columns from $H$ contains the row $(\beta_1, \ldots, \beta_{i-1}, \sigma_i + 1, \beta_{i+1}, \ldots, \beta_r)$ where $\beta_t \preceq \sigma_t$ for $t \neq i$, $t \in \{1, 2, \ldots, r\}$.

Consider the element $x = (x_1, \ldots, x_n) \in P$ in which the component $x_{j_i} = \sigma_i$ ($i \in \{1, 2, \ldots, r\}$) is not the greatest element in $P_{j_i}$ and each of the other components $x_j$ ($j \in \{1, 2, \ldots, n\} \setminus \{j_1, \ldots, j_r\}$) is the greatest element in $P_j$. Define $\sigma = (\sigma_1, \ldots, \sigma_r)$. We have the following result.

**Theorem 1.** The element $x$ is a maximal independent of $R$ element if and only if the set of columns of the matrix $L_R$ with the indexes $j_1, \ldots, j_r$ is an ordered irredundant $\sigma$-covering of $L_R$.

By Theorem 1, the logical classification of data that are the product of finite partial orders is effectively reduced to finding ordered irredundant coverings of an integer matrix.

**Remark 1.** The concept of ordered irredundant $\sigma$-covering of the matrix $L_R$ defined in this section is a generalization of the irreducible covering of a Boolean matrix. Indeed, if every $P_i = \{0, 1\}$ and the order $0 \prec 1$ is defined on $P_i$, then the ordered irredundant $(0, \ldots, 0)$-covering of $L_R$ is an irreducible covering.

## 4 Logical Classification of Partially Ordered Data

In this section, we propose a more general statement of the logical classification problem aimed at solving problems in which each feature takes values in a finite partially ordered set of numbers.

Let $M = N_1 \times ... \times N_n$ where $N_i$ ($i \in \{1, 2, ..., n\}$) is a finite set of values of the feature $x_i$ with a partial order defined on it. We assume that each set $N_i$, $i \in \{1, 2, ..., n\}$, has the greatest element $k_i$. If there is no such an element in $N_i$, then we complete $N_i$ with such an element.

The proximity of the object $S = (a_1, ..., a_n)$ in $M$ to the el.cl. $(\sigma, H)$, where $H = \{x_{j_1}, ..., x_{j_r}\}$, $\sigma = (\sigma_1, ..., \sigma_r)$, and $\sigma_i \in N_{j_i}$ for $i = 1, 2, ..., r$, is the quantity $\hat{B}(\sigma, S, H)$ that is equal to 1 if $a_{j_i} \preccurlyeq \sigma_i$ for $i = 1, 2, ..., r$ and equal to 0, otherwise. The object $S$ is said to generate the el.cl. $(\sigma, H)$ if $\hat{B}(\sigma, S, H) = 1$.

The definitions of the correct el.cl. of class $K$, the representative el.cl. of class $K$, the covering of class $K$, and test are completely extended for the general case under considerations if $B(\sigma, S, H)$ is replaced by $\hat{B}(\sigma, S, H)$.

Let $(\sigma, H)$ be an el.cl. in which $H = \{x_{j_1}, ..., x_{j_r}\}$, $\sigma = (\sigma_1, ..., \sigma_r)$, $\sigma_i \in N_{j_i}$ for $i = 1, 2, ..., r$. We assign to the el.cl. $(\sigma, H)$ the set $S_{(\sigma, H)} = (\gamma_1, ..., \gamma_n)$ from $M = N_1 \times ... \times N_n$ in which $\gamma_t = \sigma_i$ for $t = j_i$ ($i = 1, 2, ..., r$) and $\gamma_t = k_t$ for $t \notin \{j_1, ..., j_r\}$.

A covering $(\sigma, H)$ of the class $K$ is said to be irredundant if every el.cl. $(\sigma', H')$ such that $S_{(\sigma, H)} \prec S_{(\sigma', H')}$ is not a covering of class $K$. A representative el.cl. $(\sigma, H)$ for the class $K$ is said to be irredundant if every el.cl. $(\sigma', H')$ such that $S_{(\sigma, H)} \prec S_{(\sigma', H')}$ is not representative for the class $K$.

By $R(K)$ we denote the set of training objects in the class .

**Proposition 1.** The covering $(\sigma, H)$ of the class $K$ is an irredundant covering of $K$ if and only if $S_{(\sigma, H)} \in I(R(K)^+)$.

Let $\bar{K} = M \setminus K$. We will consider $\bar{K}$ as a separate class; i.e., we assume that there are only two classes $K$ and $\bar{K}$. The following result is obvious.

**Proposition 2.** The el.cl. $(\sigma, H)$ is an irredundant representative el.cl. for the class $K$ if and only if $S_{(\sigma, H)} \in I(R(\bar{K})^+)$ and $S_{(\sigma, H)} \in R(K)^+$.

Propositions 1 and 2 imply that, in the case of partially ordered data, the construction of logical classifiers based on construction of a set of irredundant coverings of a class or a set of irredundant representative el.cl. for a class requires the construction of maximal independent elements of the product of partial orders.

Note that the existence of representative el.cl. for the class $K$ is not guaranteed in the general case, and for a classification algorithm (based on construction of a set of representative el.cl.) to be correct, it is necessary that the descriptions of objects belonging to different classes are incomparable.

We show that, in the case of nonoverlapping classes, there exists a transformation of the feature description of the set $M$ as a result of which a nonempty set of irredundant representative el.cl. is formed for each class $K$ and every object in $K$ generates at least one el.cl. in this set.

Denote by $\tilde{P}$ the set coinciding the set $P$ with the reversed order relation; i.e., $x \preccurlyeq y$ in $P$ if and only if $y \preccurlyeq x$ in $\tilde{P}$.

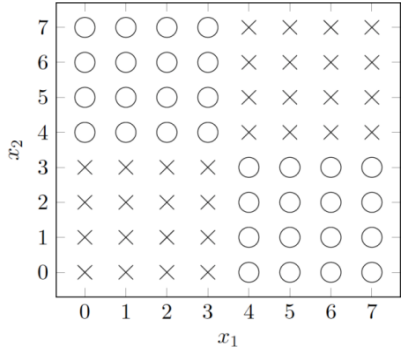 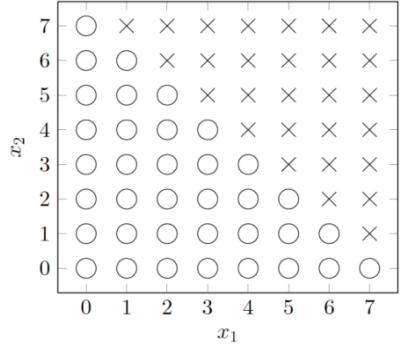

Fig. 2    Fig. 2

Let $\widetilde{M} = \widetilde{N}_1 \times \ldots \times \widetilde{N}_n$. Define the mapping $\varphi: M \to M \times \widetilde{M}$ as follows. The mapping $\varphi$ takes the object $S = (a_1, \ldots, a_n)$ in $M$ to the object $\varphi(S) = (a_1, \ldots, a_n, a_{n+1}, \ldots, a_{2n})$ in $M \times \widetilde{M}$ where $a_{i+n} = a_i$ for $i \in \{1, 2, \ldots, n\}$. In other words, the feature description of $S$ is duplicated with the reversed order relation.

Let $\varphi(A)$ ($A \subset M$) be the image of $A$ under the mapping $\varphi$. We have the following theorem.

**Theorem 2.** If the classes of the set $M$ do not overlap, then every precedent of the class $\varphi(K)$ generates an irredundant representative el.cl. of the class $\varphi(K)$.

**Remark 2.** If each set $N_i$ ($i \in \{1, 2, \ldots, n\}$) is an antichain with the added greatest element, then $R(K)^+ = R(K)$, $\hat{B}(\sigma, S, H) = B(\sigma, S, H)$, and therefore, we have in this case the classical statement considered in Section 1.

Figures 1 and 2 show illustrative examples that demonstrate the advantage of the more general statement of the logical classification problem. We consider two classification problems each of which has two classes and the system of features $\{x_1, x_2\}$. The training objects of the classes $K_1$ and $K_2$ are shown by crosses and circles, respectively. The set of possible values of each feature is the set of integers.

Consider the first example (see Fig. 1). It is easy to verify that, for every value of feature $x_1$, the number of objects in each class for which $x_1$ takes this value is four. The same is true for feature $x_2$. It is also easy to verify that in the classical case each el.cl. of rank 2 is correct either for the class $K_1$ or for $K_2$ and that it is generated by only one object in the training set. Therefore, every correct el.cl. of any of the two classes has low informativeness. Such an el.cl. is not generated by any of the objects not included in the training set. The situation in the second example (see Fig. 2) is similar. There are only two correct el.cl. of rank 1—these are the el.cl. $((0), \{x_1\})$ and $((0), \{x_2\})$; each of them is generated by only few objects of the class $K_2$.

On the other hand, if we define the natural order ($\cdots \leqslant 0 \leqslant 1 \leqslant \cdots$) on the set of possible values of the features $x_1$ and $x_2$, then the decision rules for both examples become quite simple. Let $S = (a_1, a_2)$ be an object to be classified. Then, the decision rule for the first example is as follows: if $(a_1, a_2) \leqslant (3, 3)$ or $(a_1, a_2) \geqslant (4, 4)$, then $S \in K_1$; otherwise, $S \in K_2$. The decision rule for the second example is as follows: if $(a_1, a_2) \preccurlyeq (x, 7 - x)$ for at least one $x \in \{0, 1, \ldots, 7\}$, then $S \in K_2$; otherwise, $S \in K_1$.

## 5      Experiments with Real-Life Data

The quality of classifiers was experimentally compared on five datasets taken from www.kaggle.com and archive.ics.uci.edu. For testing, we used the triple cross validation procedure, and the quality was assessed by the fraction of correct classifications.

The results are presented in Table 1. The first column shows the name of the dataset and its parameters (the number of objects $m$, the number of features $n$, the number of classes $l$, and the maximum number of values of feature $k$). Each of the following four columns corresponds to one variant of ordering the feature values in the algorithms in which the voters are irredundant representative el.cl. In addition, the classical logical corrector model developed in [Djukova et. al, 2017a] was used. This model uses a preliminary selection of informative el.cl. and boosting in learning.

**Table 1.**

| Dataset name $(m, n, l, k)$ | All features are antichains | All features are chains | Mixed features (chains and antichains) | All features are chains (with duplication of features) | Logical corrector |
|---|---|---|---|---|---|
| Car (1728, 4, 4, 4) | 73% | 70% | 84% | 81% | **97%** |
| Heart (302, 13, 2, 151) | 76% | 74% | **81%** | - | 80% |
| Ph (427, 3, 15, 256) | 43% | 10% | 51% | **63%** | 31% |
| Dermatology (336, 34, 4, 75) | **95%** | 82% | **95%** | - | 51% |
| Turkey (129, 14, 6, 6) | 35% | 30% | 35% | 39% | **42%** |

These results confirm that the classical approach (each feature is an antichain) does not take into account the dependences in the values of attributes; for this reason, this approach tends to overfitting. The linear order (chain) is very sensitive to the choice of linear order ($k!$ variants of chains for a feature with $k$ admissible values are possible). The choice of which feature is a chain and which is an antichain is ambiguous. The practice showed that features with a large number of admissible values that have a natural linear ordering (e.g., age) are good candidates for the role of chains. The presence of several chains among antichains (the version of mixed features) in the majority of problems improved the quality and reduced the computation time. For the datasets Car, Ph and Turkey, the heuristic with duplicating features in reversed order was also considered (see Theorem 2). It is seen that when all features are chains, duplication yields better results than in the version without duplication. For the datasets Heart, Ph, and Dermatology, which have a large number of feature values, the new algorithms demonstrated higher quality compared with logical corrector.

# 6      Conclusions

For the supervised classification problem (machine learning), the logical analysis of data represented by the product of partially ordered finite sets (product of partial orders) is studied for the first time. Based on a generalization of the basic concepts, the conventional approach to constructing logical classification procedures is improved. Results of testing the new classification algorithms on model and real-life data are presented. This study has an important scientific methodological significance and considerably extends the field of practical application of the logical data analysis methods.